\documentstyle[times,twocolumn,epsfig,aps,prb]{revtex}

\title{Many--particle entanglement with Bose--Einstein condensates}

\author{A.~S\o rensen$^*$, L.-M.~Duan$^\dagger$,
  J.~I.~Cirac$^\dagger$ \& P.~Zoller$^\dagger$}
\address{$^*$ Institute of Physics and Astronomy, University of
  Aarhus, DK-8000 \AA rhus C, Denmark\\
  $^\dagger$ Institute for Theoretical Physics, University of
  Innsbruck, A-6020 Innsbruck, Austria}

\begin{document}
%\twocolumn
\maketitle
\begin{abstract} 
We propose a method to produce entangled states of several particles
starting from a 
Bose-Einstein condensate. In the proposal, a single fast $\pi/2$ pulse is
applied to the atoms and due
to the collisional interaction, the subsequent free time evolution 
creates an entangled state involving  all atoms in the condensate.
The created entangled state is a spin-squeezed state which could be
used to improve the sensitivity of atomic clocks. 
\end{abstract}

%\small
The possibility of creating and manipulating entangled states of
many--particle systems has recently boosted the field of quantum
information since it may yield new applications which rely on
the basic principles of Quantum Mechanics \cite{PhysWorld}.
So far, up to four atoms have been
entangled in a controlled way \cite{sackett,Haroche}.
The
experimental achievement of Bose--Einstein condensation
\cite{Cornell,Ketterle,General} has also raised a lot of attention
since it may lead to  applications in several fields of
Science. Some of these applications are based on the fact that
condensates can be considered as pure states at the single
particle level, which is a crucial requirement for the production of
entangled state.  Thus, a natural question is to investigate
whether Bose--Einstein condensates can also be used in some
applications of quantum information. In this paper we show that
this is indeed the case. We propose a method to obtain
substantial entanglement of a large number of atoms with 
present technology. In our proposal, a single resonant pulse is applied to
all atoms in 
the condensate, and the collisional interaction entangles the
atoms in the subsequent free evolution.

% , with
% such a degree of entanglement that one could reduce the
% projection noise in atomic clocks by several orders of
% magnitude.

Consider a set of $N$ two--level atoms 
confined by some external trap. In order to describe the
internal properties of these atoms, it is convenient to let the internal
states $|a\rangle_n$ and $|b\rangle_n$
of the $n$-th atom represent the two states of a fictitious
spin 1/2 particle with angular momentum operators
$j^{(n)}_z=1/2(|a\rangle\langle a|_n - |b\rangle\langle b|_n)$,
$j^{(n)}_x=1/2(|b\rangle\langle a|_n + |a\rangle\langle b|_n)$, and
$j^{(n)}_y=i/2(|b\rangle\langle a|_n - |a\rangle\langle b|_n)$. We
consider collective effects of the atoms which are described by
total  angular momentum
operators, $\vec J=\sum_{n=1}^N \vec j^{(n)}$.
% , where $\vec
% j^{(n)}$ is the angular momentum operator corresponding to the
% $n$--th atom.
The entanglement properties of the atoms can be
expressed in terms of the variances and expectation values of
these operators. In the appendix we show
that if
\begin{equation}
\label{xi}
\xi^2 \equiv \frac{N (\Delta J_{\vec n_1})^2}
{\langle J_{\vec n_2} \rangle^2 +
\langle J_{\vec n_3}\rangle^2} < 1,
\end{equation}
where $J_{\vec n}\equiv \vec n\cdot \vec J$ and the $\vec n$'s
are mutually orthogonal unit vectors, then the state of the
atoms is non--separable (i.e. entangled).
% A proof of Eq.~(\ref{xi}) is given
%at the end of the present letter. 
The
parameter $\xi^2$ thus characterizes the atomic entanglement, and
states with $\xi^2 < 1$ are often referred to as
``spin squeezed states'' \cite{kitagawa}.
In the following we
show how to reduce $\xi^2$ by several orders of magnitude
using  the collisional
interactions between atoms in a Bose--Einstein condensate.

We consider  a two
component weakly interacting Bose-Einstein condensate, which is present in
several
laboratories\cite{hall,stenger}, and we assume that
the interactions do not change the internal
state of the atoms.
%(below we show how to achieve this in practice). 
This
situation is described by the second quantized Hamiltonian
\begin{eqnarray}
H&=&\sum_{j=a,b} \int d^3 r \hat{\Psi}_j^\dagger({\bf r}) H_{0,j}
\hat{\Psi}_j({\bf r}) \nonumber\\
 &+&\frac{1}{2} \sum_{j=a,b} U_{jj} \int d^3r \hat{\Psi}_j^\dagger({\bf
   r})  \hat{\Psi}_j^\dagger({\bf r})  \hat{\Psi}_j({\bf r})
  \hat{\Psi}_j({\bf r}) \nonumber \\
 &+&U_{ab}\int d^3 r  \hat{\Psi}_a^\dagger({\bf
   r})  \hat{\Psi}_b^\dagger({\bf r})  \hat{\Psi}_a({\bf r})
  \hat{\Psi}_b({\bf r}),
\label{hamilton}
\end{eqnarray}
where $H_{0,j}$ is the one particle Hamiltonian for atoms in
state $j$ including the kinetic energy and the external trapping
potential $V_j({\bf r})$, $\hat{\Psi}_j({\bf r})$ is the field
operator for atoms in the state $j$, $U_{jk}=4\pi \hbar^2
a_{jk}/m$ is the strength of the interaction between particles
of type $j$ and $k$, parameterized by the scattering length
$a_{jk}$, and $m$ is the atomic mass.

Assume that we start with a Bose--Einstein condensate in state
$|a\rangle$ at very low temperature ($T\simeq 0$), so that all
the atoms are in a single particle (motional) state
$|\phi_0\rangle$. A fast $\pi/2$ pulse between the states
$|a\rangle$ and $|b\rangle$ prepares the atoms in the state
$|\phi_0\rangle^{\otimes N}\otimes(|a\rangle+|b\rangle)^{\otimes
N}/2^{N/2}$ which is an eigenstate of the $J_x$ operator with
eigenvalue $N/2$. If we choose in (\ref{xi}) $\vec
n_1=[0,\cos(\theta),\sin(\theta)]$ and $\vec n_2$ along the $x$
axis of the fictitious angular momentum, we have $\xi_\theta^2=1$
at $t=0$. By using the equations of motion of the angular
momentum operators in the Heisenberg picture we find the time
derivative of $\xi_\theta^2$ at $t=0$
\begin{equation}
\frac{d}{dt} \xi_\theta^2=\sin(2\theta)
 \frac{(N-1){\left(U_{aa}+U_{bb}-2U_{ab}\right)}}{2\hbar}
 \int d^3 r {\left| \phi_0 \right|}^4.
\label{derivative}
\end{equation}
This equation immediately shows that spin squeezing will be
produced for certain angles $\theta$ if $U_{aa}+U_{bb}\neq
2U_{ab}$ as it is in the Na experiments
\cite{stenger}. 

%Having seen that spin squeezing can be produced with
%Bose--Einstein condensates, let us
To quantify the amount of
squeezing which may be obtained,  we  assume identical trapping potentials
$V_a({\bf r})=V_b({\bf
r})$ and identical coupling constants for interaction between
atoms in the same internal state $a_{aa}=a_{bb}> a_{ab}$.
%, for the sake of
%simplicity.
Physically, this could correspond to the $|F=1,M_F=\pm 1\rangle$
hyperfine states of Na trapped in an optical dipole trap. Due to
the symmetry of these states their scattering lengths and
trapping potentials will be identical; moreover, due to angular
momentum conservation there are no spin exchanging collisions
between these states as required by our model. This is exactly
the experimental setup used in Ref. \onlinecite{stenger}. In
this experiment it is shown that these states have an
antiferromagnetic interaction $a_{ab}<a_{aa}$ which according to
(\ref{derivative}) enables the production of squeezed states.
To avoid spin changing collisions that populate
the state $|M_F=0\rangle$\cite{miesner} one has to slightly modify such an
experimental
set--up.
If the $F=1$ manifold is coupled to the
$F=2$ with a far off-resonant blue detuned $\pi$-polarized
microwave field, the
$|F=1,M_F=0\rangle$ state is raised in energy with respect to
the $|F=1,M_F=\pm 1\rangle$ states, since the Clebsch-Gordan coefficient
for the
$|F=1,M_F=0\rangle \rightarrow |F=2,M_F=0\rangle$ is larger than
the coefficients for the $|F=1,M_F=\pm 1\rangle\rightarrow
|F=2,M_F= \pm 1\rangle $ transitions, and spin exchanging
collisions become energetically forbidden. If for instance one chooses a
detuning $\delta=(2\pi)25$ MHz and a resonant Rabi frequency for the
$1\rightarrow 1$ transition of $\Omega =(2\pi) 2 $ MHz, the energy
separation is $\Delta E=640$nK. 
With a  typical chemical potential   
$\mu\approx220$ nK\cite{miesner}, this energy separation is much higher
than the available energy in the collisions and the $|F=1,M=0\rangle$ state
is completely 
decoupled. In Ref. \onlinecite{miesner} a much smaller
energy difference is shown  to exclude spin exchanging collision and we
therefore expect that much weaker fields will suffice. 

The assumption $a_{aa}=a_{bb}$ has several advantages.
%\cite{foot3}.
Firstly, it reduces the effect of fluctuations in the total
particle number. If $a_{aa} \neq a_{bb}$ the mean spin performs
a $N$ dependent rotation around the $z$-axis, and fluctuations
in the number of particles introduces an uncertainty in the
direction of the spin which effectively reduces the average
value and introduces noise into the system. With $a_{aa}=a_{bb}$
the mean spin remains in the $x$-direction independent of the
number of atoms in the trap. Secondly, this condition ensures a
large spatial overlap of different components of the
wavefunction. After the $\pi/2$ pulse the spatial wavefunction
is no longer in the equilibrium state. That is, due to the
atomic repulsions (which are now different than before since the
atoms are in different internal states), the spatial
distribution of the atomic cloud will start oscillating.
Furthermore, since the state of the system is now distributed
over a range of number of particles in the $|a\rangle$ state
($N_a$), and since this number enters into the time evolution,
the $N_a$ dependent wavefunctions $\phi_a$ and $\phi_b$ are
different for particles in the states $|a\rangle$ and
$|b\rangle$. With $a_{aa}=a_{bb}$, $\phi_a$ and $\phi_b$ are
identical if $N_a$ equals the average number $N/2$. In the limit
of large $N$, the width of the distribution on different $N_a$'s
is much smaller than $N_a$ and all the spatial wavefunctions are
approximately identical $\phi_a (N_a,t)\approx\phi_b
(N_a,t)\approx \phi_0 (t)$. This relation is only true if
$a_{aa}>a_{ab}$ where small deviations from the average
wavefunction perform small oscillations. In the opposite case
the deviations grow exponentially \cite{sinatra} resulting in a
reduction of the overlap of the $a$ and the $b$ wavefunctions
and a reduced squeezing. The advantages mentioned above could
also be achieved with $a_{aa}\ne a_{bb}$ by using the
breath--together solutions proposed in Ref.
\onlinecite{sinatra}.
% , but here we
% only consider $a_{aa}= a_{bb}$.

Before analyzing quantitatively the complete system, we
estimate the amount of spin squeezing we can reach with our
proposal by using a simple model. Assuming the same wavefunction
$\phi_0$ for both $|a\rangle$ and $|b\rangle$ atoms is constant
and independent of the number $N_a$, the spin dependent part of
the Hamiltonian~(\ref{hamilton}) may be written as $H_{\rm
spin}=\hbar \chi J_z^2$, where $\chi$ depends on the scattering lengths
and the wavefunction $\phi_0$. The spin squeezing produced by
this Hamiltonian can be calculated exactly \cite{kitagawa}. In
the limit of large $N$, the minimum obtainable squeezing
parameter is
$\xi_\theta^2=\frac{1}{2}{\left(\frac{3}{N}\right)}^{2/3}$,
which indicates that our proposal might produce a reduction of
$\xi^2$ by a factor of $\sim N^{2/3}$ which would be more than
three orders of magnitude with $10^5$ atoms in the condensate.

\begin{figure}[htbp]
  \begin{center}
    \epsfig{width=7cm,file=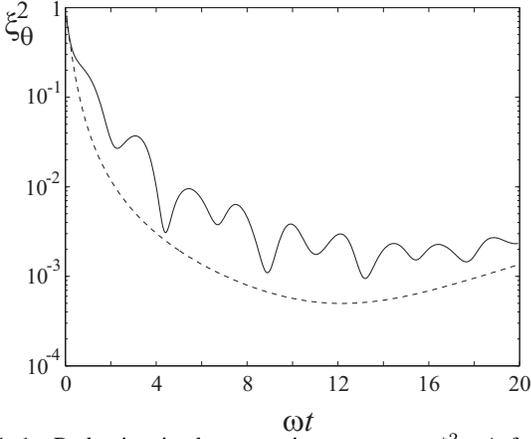}
    \caption[]{Reduction in the squeezing parameter $\xi^2$. A fast
      $\pi/2$ pulse between two internal states is applied to all
      atoms in the condensate. The subsequent free time evolution results
      in a strong squeezing of the total spin. The angle $\theta$ is
      chosen such that $\xi_\theta^2$ is minimal. The solid line is the
      results
      of a  numerical integration (see text).  For numerical convenience we have assumed a spherically
symmetric potential $V(r)=m\omega^2 r^2/2$. The  parameters are
      $a_{aa}/d_0=6*10^{-3}$, $a_{bb}=2a_{ab}=a_{aa}$, and $N=10^5$.
      The dashed curve shows the
      squeezing obtained from the Hamiltonian $H_{\rm spin}=\hbar \chi
      J_z^2$. The parameter
      $\chi$ is chosen such  that the reduction of $J_x$ obtained from the
      solution in Ref.~\onlinecite{kitagawa} is consistent with the results
      of Ref.~\onlinecite{sinatra}.}
    \label{fig:reduction}
  \end{center}
\end{figure}

In contrast to the simplified Hamiltonian $H_{\rm spin}$ the real
Hamiltonian (\ref{hamilton}) will also entangle the internal and motional
states of the atoms which is a source of decoherence for the spin
squeezing. To quantify this effect we 
have performed a direct numerical integration following the procedure
developed in Ref. \onlinecite{sinatra}. We split the whole Hilbert space
into orthogonal subspaces containing a fixed number of particles $N_a$ and
$N_b=N-N_a$ in each of the internal states, respectively.  In each subspace
we make a Hartree-Fock variational ansatz in terms of three-dimensional
spatial wavefunctions $\phi_a(N_a,t)$ and $\phi_b(N_b,t)$, which are
evolved according to 
the time dependent coupled Gross-Pitaevskii
equations. This is an
approximation to the full problem which is  valid in 
the limit of low temperatures and short times, where the population of the
Bogoliubov modes is small.  Particularly it is a good
approximation in our case with $a_{ab}< a_{aa}$, $a_{bb}$ where there are
no demixing instabilities. With this procedure, the decoherence induced by
the entanglement with the  
motional state is effectively taken into account. 
Together with the prediction from the simple Hamiltonian $H_{\rm
spin}$ the result of the simulation is shown in
Fig.~\ref{fig:reduction}.  The two curves are roughly in
agreement confirming that the system is able to approximate the
results of the Hamiltonian  $H_{\rm
spin}$. The numerical solution shows
fluctuations due to the oscillations of the spatial
wavefunction. The large dips at $\omega t\approx$ 4, 9, 13, and
18 are the points where the atomic cloud reach the
initial width. At these instants the overlap of the
wavefunctions is maximal and the two curves are very close (up
to a factor of two). The small dips at $\omega t=$2, 7, 11, and
16 corresponds to the points of maximum compression. With the
realistic parameters used in the figure, our simulation suggests
that three orders of magnitude squeezing is possible. Also, note
the time scale in the figure. The maximally squeezed state is
reached after approximately two oscillation periods in the trap.
For a fixed ratio of the scattering lengths $a_{ab}/a_{aa}$, the
optimal time scales as $(a_{aa}/d_0)^{-2/5}N^{-1/15}$, where
$d_0=\sqrt{\hbar/(m\omega)}$ is the width of the ground state of
the harmonic potential.

\begin{figure}[b]
  \begin{center}
    \epsfig{width=7cm,file=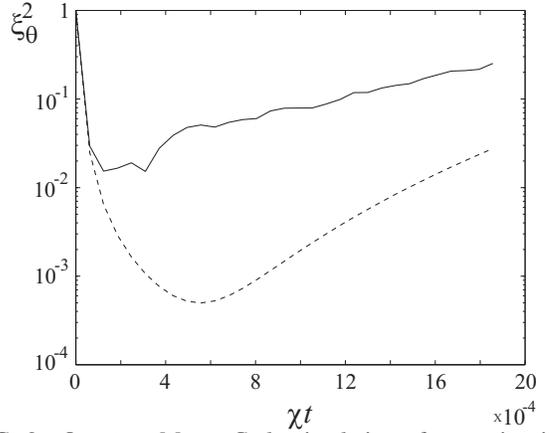}
    \caption{Quantum Monte Carlo simulation of squeezing in the
      presence of loss. The full line shows squeezing obtained from
      the Hamiltonian
      $H_{\rm spin}$ with particle loss described by a constant
      loss rate $\Gamma=200\chi$. The dashed curve shows squeezing
      without particle loss.}
    \label{fig:loss}
  \end{center}
\end{figure}

The analysis so far has left out a number of possible
imperfections. Specifically, we have assumed that all scattering
length are real so that no atoms are lost from the trap and we
have not considered the role of thermal particles. To estimate
the effect particle losses we have performed a Monte Carlo
simulation\cite{klaus} of the evolution of squeezing from the
Hamiltonian $H_{\rm spin}$. The particle loss is
phenomenologically taken into account by introducing a loss rate
$\Gamma$ which is identical for atoms in the $|a\rangle$ and the
$|b\rangle$ state. In Fig.~\ref{fig:loss} we show the
obtainable squeezing in the presence of loss. Approximately 10
\% of the atoms are lost at the time $\chi t\approx 6*10^{-4}$
where the squeezing is maximally without loss. With the
parameters of Fig.~\ref{fig:reduction} this time corresponds to
roughly two trapping periods. Such a large loss is an
exaggeration of the loss compared to current experiments and the
simulation indicate that even under these conditions, squeezing
of nearly two orders of magnitude may be obtained. On the other
hand, the effects of thermal particles can be suppressed at
sufficiently low temperatures but due to the robustness with
respect to particle loses shown in Fig.\ 2, we expect to obtain
high squeezing even at some finite temperatures.
% In the numerical analysis
%we used the simplifying assumption that the spatial wavefunction for each
%value of $N_a$ could be described by the mean field solution of the
%Gross-Pitaevskii equation. 

In conclusion we believe that we have presented a simple and robust
method to
produce entangled states of a large number of atoms with present
technology. The produced entangled state are interesting in fundamental
physics and they also have possible technological applications in atomic
clocks \cite{wineland,Kasevich}, where the projection noise $(\Delta J_\theta)^2$ is currently the main
source of noise \cite{santarelli}.  In particular Boyer and Kasevich \cite{Kasevich} have shown how to use an entangled state with half of the particles $N/2$ in one internal state $a$, and the other half in the other state $b$ to obtain the Heisenberg limit in atomic interferometric measurements. On the other hand, it has been shown that if the atoms are
prepared in a state with  $\xi^2<1$ before they are injected into an atomic
clock, one can reduce the frequency
noise (variance in the frequency measurements) or the  measuring
  time to obtain a desired precision  by
a factor $\xi^2$ as compared to the case in which one uses atoms
in an uncorrelated state \cite{wineland}. If the squeezing
produced by our proposal is transferred into a suitable clock
transition, and the atomic cloud is allowed to expand so that
the role of collisions is reduced during the frequency
measurement, the states could be directly applied in the current
set--up of atomic clocks. Other theoretical proposals for noise
reduction in neutral atom systems have been made
 \cite{Sorensen}, and a weak squeezing of the spin has
 recently been produced experimentally \cite{hald,kuzmich}. 
However, our proposal has the considerable advantage that it offers a very
strong noise reduction, and it is directly applicable to existing
experimental set--ups.
In future experiments with negligible
particle loss even for very long interaction times, the Hamiltonian
$H_{\rm spin}$ could also be used to produce  maximally
entangled state of any number of atoms \cite{sackett,klausCat}, which could
reduce the frequency noise to the fundamental limit of quantum
mechanics \cite{Bollinger}.

\noindent {\bf Appendix}\\
%{\scriptsize
Here we present the derivation of
Eq.~(\ref{xi}) as a criterion for entanglement. An $N$-particle density 
matrix $\rho$ is defined to be separable 
(non-entangled) if it can be decomposed into 
\begin{equation}
  \rho=\sum_k P_k \rho_k^{(1)}\otimes  \rho_k^{(2)} \otimes
  ... \otimes  \rho_k^{(N)},
\label{separable}
\end{equation}
where the coefficients $P_k$ are positive real
numbers fulfilling $\sum_k
P_k=1$, and  
$\rho_k^{(i)}$ is a density matrix for the $i$'th particle.
The variance of $J_z$ may be described as $(\Delta
J_z)^2=\frac{N}{4}-\sum_k P_k\sum_i  \langle 
 j_z^{(i)}\rangle_k ^2+\sum_k P_k \langle J_z
 \rangle_k^2 -\langle J_z \rangle^2$, and using  Cauchy-Schwarz's
 inequality  and
 $\langle j_x^{(i)}\rangle_k^2+ \langle
  j_y^{(i)}\rangle_k^2+ \langle j_z^{(i)}\rangle_k^2\leq 1/4$ 
   we
  find three inequalities for separable states $\sum_k P_k \langle J_z 
  \rangle_k^2 \geq 
  \langle J_z \rangle^2$, $-\sum_k P_k \sum_i \langle
    j_z^{(i)}\rangle_k^2 
    \geq -\frac{N}{4} +\sum_k P_k \sum_i \langle
    j_x^{(i)}\rangle_k^2+\langle
    j_y^{(i)}\rangle_k^2$, and 
     $\langle J_x\rangle^2\leq 
      N \sum_k P_k \sum_i \langle
      j_x^{(i)}\rangle_k^2$. From these inequalities we
      immediately
      find that any separable state obeys $\xi^2\geq 1$ and hence any
      state with $\xi^2<1$ is non-separable.%}

\noindent {\bf Acknowledgments}

\noindent
This work was supported by the Austrian Science Foundation, the
European Union project EQUIP, the TMR
european network, the ESF under the PESC
program "Quantum Information", the Institute for Quantum
Information GmbH, and Thomas B. Thriges Center for Kvanteinformatik.
A. S. acknowledges the hospitality of the University of Innsbruck.

\end{document}